\documentclass[twoside,a4paper,11pt]{sea22}
\usepackage{graphicx}
\usepackage{hyperref}
\usepackage{media9}
\begin{document}
\pagenumbering{arabic}
\pagestyle{myheadings}
\thispagestyle{empty}
{\flushleft\includegraphics[width=\textwidth,bb=58 650 590 680]{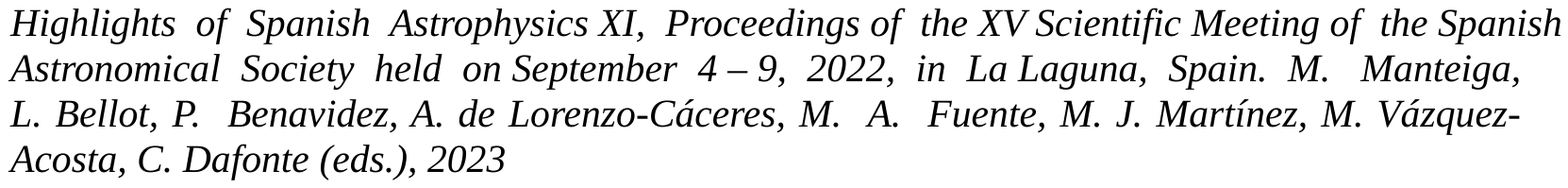}}
\vspace*{0.2cm}
\begin{flushleft}
{\bf {\LARGE
%
Parameters for $>300$ million {\it Gaia} stars:\\
Bayesian inference vs. machine learning
%
}\\
\vspace*{1cm}
%
Anders, F.$^{1}$,
Khalatyan, A.$^{2}$, 
Queiroz, A.$^{2,3}$, 
Nepal, S.$^{2}$, and
Chiappini, C.$^{2}$
%
}\\
\vspace*{0.5cm}
%
$^{1}$
Institut de Ciències del Cosmos (IEEC-UB), Dept. Física Quàntica i Astrofísica (FQA), Universitat de Barcelona, C Martí i Franqués, 1, 08028 Barcelona, Spain\\
$^{2}$
Leibniz-Institut f\"ur Astrophysik Potsdam (AIP), An der Sternwarte 16, 14482 Potsdam, Germany\\
$^{3}$
Institut f\"{u}r Physik und Astronomie, Universit\"{a}t Potsdam, Haus 28 Karl-Liebknecht-Str. 24/25, D-14476 Golm, Germany
%
\end{flushleft}
%
\markboth{
Bayesian inference vs. Machine learning for stellar parameters
}{ 
%
Anders et al.
%
}
\thispagestyle{empty}
\vspace*{0.4cm}
\begin{minipage}[l]{0.09\textwidth}
\ 
\end{minipage}
\begin{minipage}[r]{0.9\textwidth}
\vspace{1cm}
\section*{Abstract}{\small
%
The {\it Gaia} Data Release 3 (DR3), published in June 2022, delivers a diverse set of astrometric, photometric, and spectroscopic measurements for more than a billion stars. The wealth and complexity of the data makes traditional approaches for estimating stellar parameters for the full {\it Gaia} dataset almost prohibitive.
We have explored different supervised learning methods for extracting basic stellar parameters as well as distances and line-of-sight extinctions, given spectro-photo-astrometric data (including also the new {\it Gaia} XP spectra). For training we use an enhanced high-quality dataset compiled from {\it Gaia} DR3 and ground-based spectroscopic survey data covering the whole sky and all Galactic components. We show that even with a simple neural-network architecture or tree-based algorithm (and in the absence of {\it Gaia} XP spectra), we succeed in predicting competitive results (compared to Bayesian isochrone fitting) down to faint magnitudes.
We will present a new {\it Gaia} DR3 stellar-parameter catalogue obtained using the currently best-performing machine-learning algorithm for tabular data, {\tt XGBoost}, in the near future.
\normalsize}
\end{minipage}
\section{Introduction \label{intro}}

The {\it Gaia} mission \cite{Gaia2016} has triggered an enormous increase in astronomical data that is revolutionising not only stellar and Galactic science but also has implications for cosmology and fundamental physics. 
The latest {\it Gaia} data release, DR3 \cite{Gaia2023}, for the first time contains also low-resolution spectra, taken with {\it Gaia}'s blue and red photometer (BP/RP), for 219 million sources. These so-called XP spectra \cite{DeAngeli2023} represent the biggest homogeneous spectroscopic dataset (albeit at very low resolution; $R\sim25$) available to date. Efficient methods to extract information from this dataset are starting to appear (e.g. \cite{Weiler2022, Andrae2023}). 

\section{Bayesian inference of stellar parameters with {\tt StarHorse} \label{sh}}

In the past years, our group has been developing an efficient Bayesian isochrone-fitting code, {\tt StarHorse} \cite{Santiago2016, Queiroz2018, Queiroz2020}, to infer stellar parameters, distances, and extinctions, to be able to analyse the ever-growing stellar spectroscopic survey datasets.

\begin{figure}[!h]
\center
\includegraphics[width=0.9\textwidth,angle=0,clip=true]{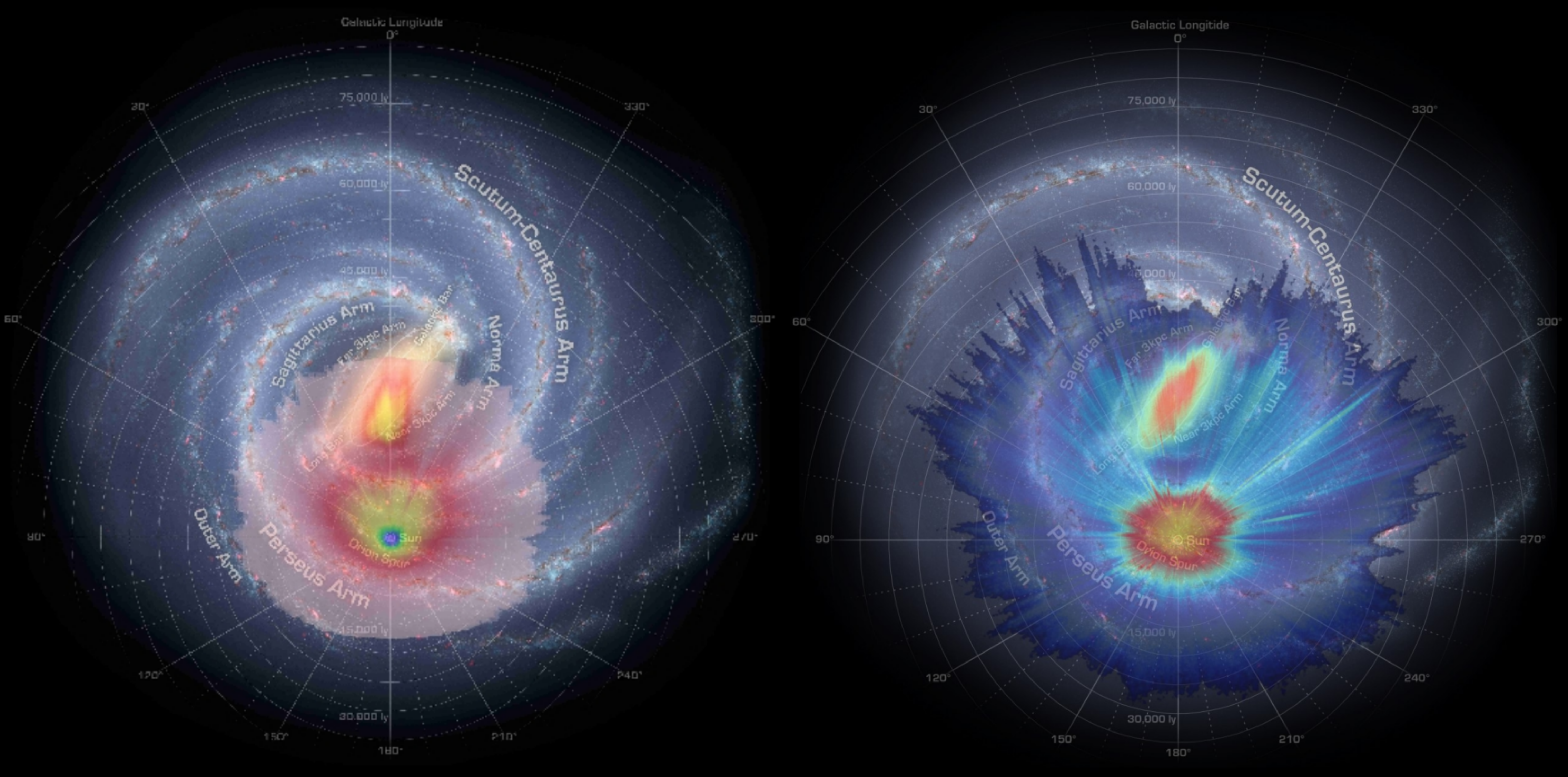} 
\caption{\label{fig1} Comparison of the 5-year mission expectation for the Galactic coverage of the {\it Gaia} data before launch (left; \cite{Luri2014}) with the results from {\tt StarHorse} for {\it Gaia} DR2 ($G<18$, 22 months of observation, right; \cite{Anders2019}).}
\end{figure}

In \cite{Anders2019}, we applied the code for the first time to the {\it Gaia} data without spectroscopic information. The first experiments proved to be so promising that a big-data
analysis run (285 million {\it Gaia} DR2 stars with $G<18$, cross-matched with the Pan-STARRS1, 2MASS, and WISE photometric catalogues) was carried out using the computing cluster of the Leibniz-Institut für Astrophysik Potsdam (AIP). Our results
doubled the number of Gaia DR2 sources with astrophysical parameters, improved the
accuracy of the geometric distances, and revealed the presence of the Galactic bar in the Gaia
data in a direct and completely unexpected manner (see Fig. \ref{fig1}).

After the release of {\it Gaia}'s Early Data Release 3 in 2021 \cite{Gaia2021}, and building on the success of the previous {\tt StarHorse} catalogue, we ran our code on Gaia EDR3 data coupled with other large-area photometric surveys (now also including SkyMapper data). This resulted in an improved catalogue of 362 million stars, down to magnitude $G<18.5$ (published in \cite{Anders2022}). Thanks to the more precise EDR3 parallaxes and drastically reduced systematics \cite{Lindegren2021}, the Galactic density maps derived from it now probe a much greater volume than in \cite{Anders2019}, extending to regions beyond the Galactic bar and to Local Group galaxies, with a larger total number density.

The code was also sped up by using a less dense stellar model grid and a new computing cluster, which improved the CO$_2$ footprint of the project by factor $\sim6$, but still consumed around 1 month of computing time on a 1000-core cluster. At latest with the upcoming LSST  surveys, this type of endeavour will therefore become unfeasible. In addition, {\tt StarHorse} is unable to process spectroscopic measurements (such as the XP spectra) directly, which is an obvious drawback given that these data contain valuable information on the stellar parameters \cite{Fouesneau2023}.  

\section{Predicting stellar parameters with a simple neural network \label{shnet}}

\begin{figure}[!h]
\center
\includegraphics[width=0.9\textwidth,angle=0,clip=true]{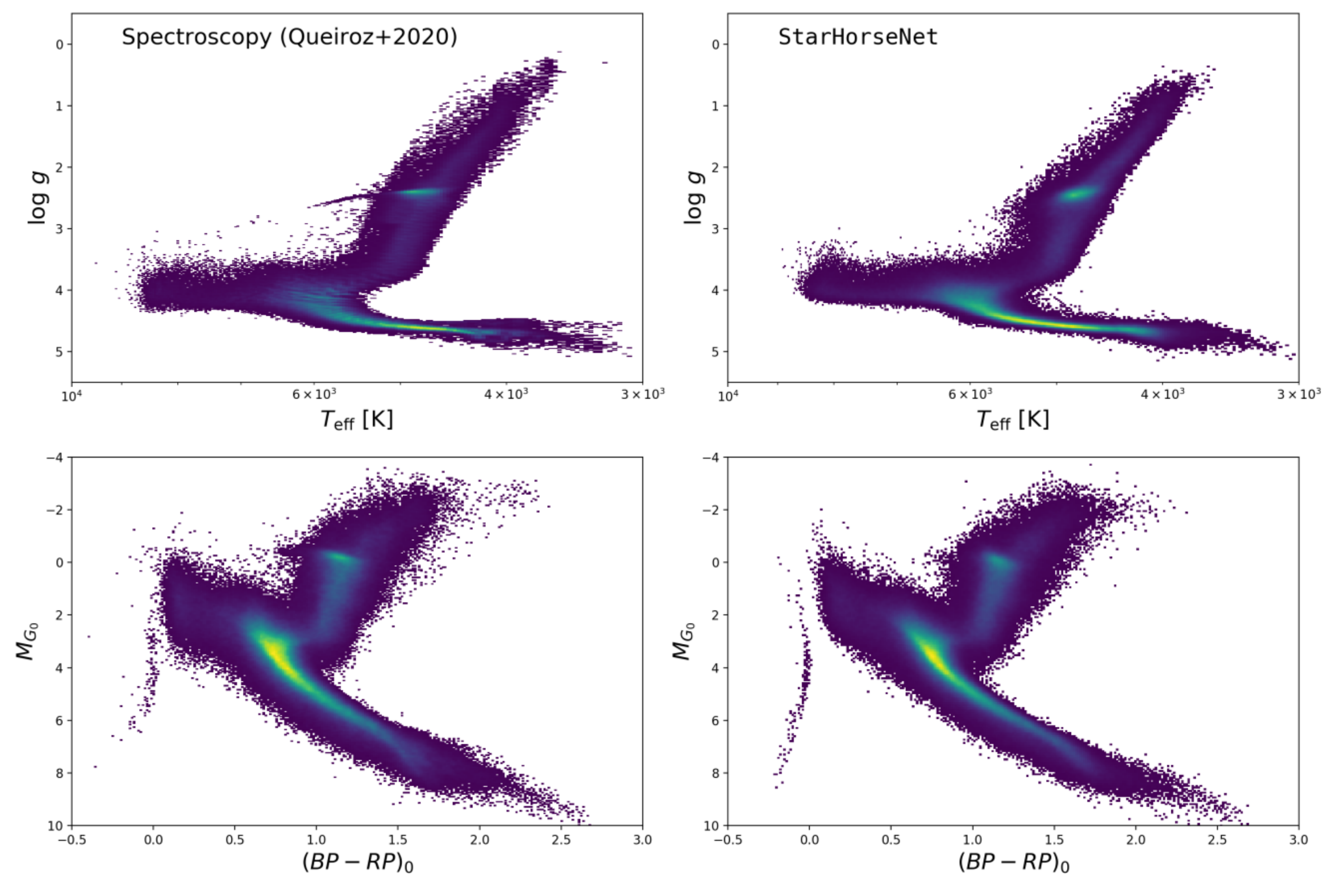} 
\caption{\label{fig2} Results of the neural-network approach to stellar parameter estimation. Comparison between the colour-magnitude and Kiel Diagrams produced by the ANN model (right) with the spectroscopic test dataset \cite{Queiroz2020}. Plot from the MSc thesis of Rany Assaad (University of Surrey, 2021).}
\end{figure}

In view of the computational drawbacks of traditional isochrone-based methods, in 2020 we started testing artificial neural networks (ANN) to predict stellar parameters for {\it Gaia} DR2 stars together with R. Assaad, an Erasmus+ MSc student visiting from the University of Surrey. As a training and test dataset, we used the {\tt StarHorse} labels for various large-scale spectroscopic stellar surveys available at the time \cite{Queiroz2020} (see Fig. \ref{fig2}).

This experiment was also surprisingly successful: with respect to \cite{Anders2019} the ANN  technique (even a very basic multi-layer perceptron architecture) doubled the number of stars (now more tha 300 million) considered to be of good quality. The full pipeline (including training and prediction) could now be run on a 48-core machine within only 3 days, and it produced competitive posterior uncertainties (using a Monte-Carlo drop-out technique). The median uncertainties amounted to 16\% in distance, 0.15 mag in $V$-band extinction $A_V$, 155 K in effective temperature $T_{\rm eff}$. 

After cleaning the results of our Gaia DR2 run for poor input and output data, a sample of 373 million converged stars remained, for which we achieve a median uncertainty of 16\% in distance, 0.15 mag in V-band extinction, and 155 K in effective temperature. Our experiment showed that even with a simple neural-network structure, one can succeed in predicting competitive results based on {\it Gaia} DR2 down to fainter magnitudes compared to classical isochrone or spectral-energy distribution fitting methods. Independent recent work \cite{Fallows2022} reached similar concluisons.

\section{Improved results with tree-based algorithms \label{shboost}}

\begin{figure}[!h]
\center
\includegraphics[width=0.9\textwidth,angle=0,clip=true]{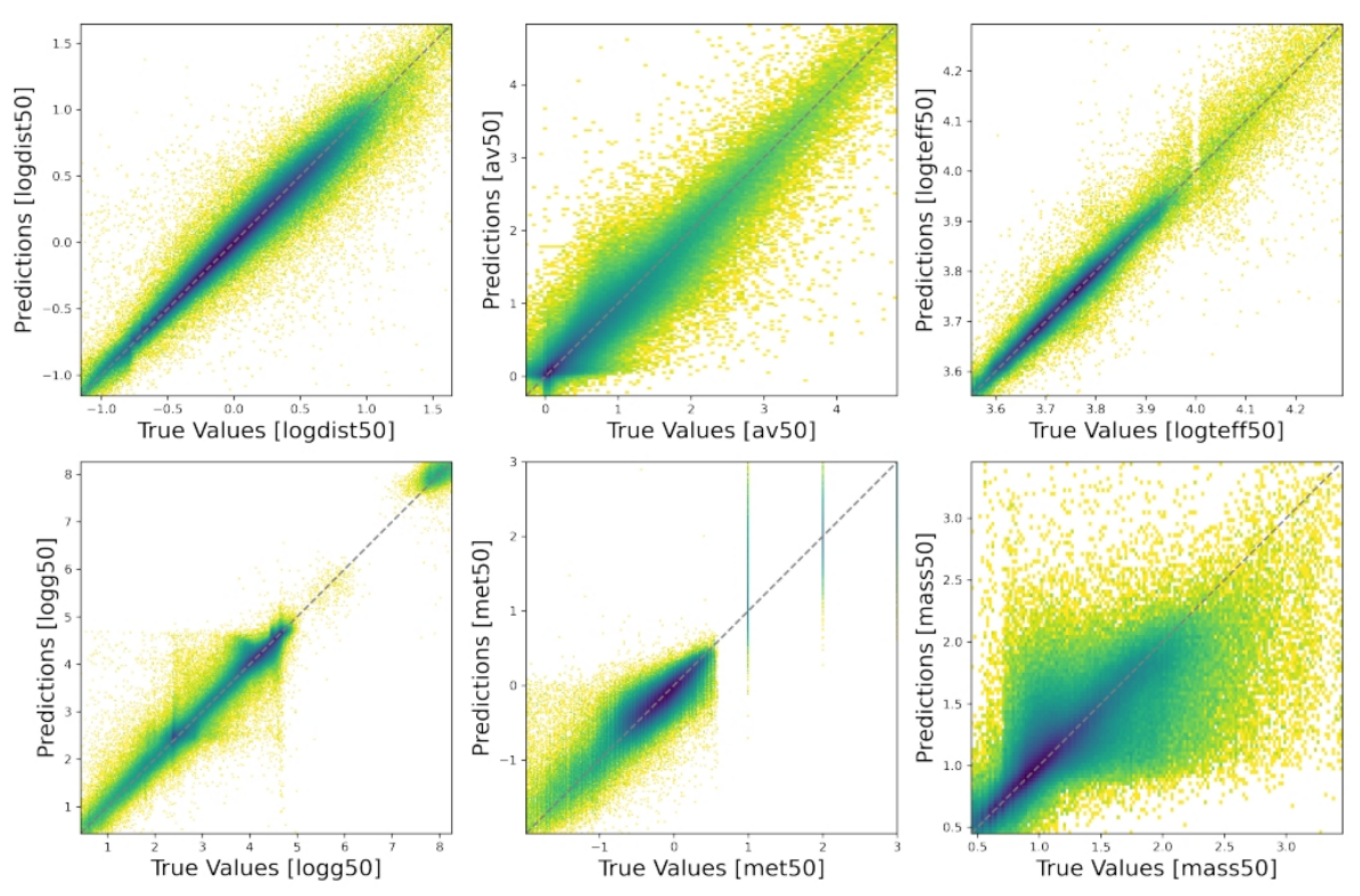} 
\caption{\label{fig3} First results (June 2022) of the {\tt XGBoost} approach to stellar parameter estimation for the case of stars with available {\it Gaia} DR3 XP spectra. One-to-one comparisons between the predicted stellar parameters and the spectroscopic test dataset \cite{Queiroz2023} (labelled "true values").}
\end{figure}

The {\it Gaia} DR3 XP spectra are not delivered in the form of classical spectra, but come in the form of internally-calibrated spectral coefficients \cite{Carrasco2021}. They can therefore be treated as tabular data and easily fed to a supervised learning technique.
\cite{Borissov2021} benchmark-tested several regression algorithms (including several neural network architectures) for tabular data and found that the most efficient and accurate technique was the tree-based algorithm Extreme Gradient-Boosted Trees ({\tt XGBoost}, \cite{Chen2016}). This technique is implemented in the {\tt python} package {\tt xgboost} that is widely used for classification tasks in astronomy (e.g. \cite{Bethapoudi2018, Yi2019, Li2021, Cunha2022}). Examples for the use of {\tt XGBoost} for regression tasks in astronomy are sparser, but also start to appear, e.g. for photometric redshifts \cite{Chong2019, Li2022}, predicting the number of sunspots \cite{Dang2022}, or determining spectroscopic stellar ages \cite{Hayden2022}.


For our first tests, we assembled a training set of 4 million {\it Gaia} DR3 stars with XP spectra that have also been observed by spectroscopic stellar surveys and have {\tt StarHorse} stellar parameters \cite{Queiroz2023}. As training columns we used the normalised XP coefficients, as well as {\it Gaia} astrometry and broad-band optical and infra-red photometry. The training set was complemented with spectroscopically observed white dwarfs from \cite{Gentile2021}. The predicted labels include distance, $A_V$, $T_{\rm eff}$, surface gravity log $g$, metallicity [M/H], and stellar mass. 

The first tests yielded the following precisions and accuracies for unseen test data: $\Delta \log d= –0.002 \pm 0.075, \Delta A_V = +0.001 \pm 0.145, \Delta \log T_{\rm eff} = 0.000 +- 0.011$ (see Fig. \ref{fig3}). The computational cost is even lower than in the case of the simple neural network (Sect. 3), and robust to both the choice of hyperparameters and the scaling of the input data. For the first time, our group is also able to deliver acceptable stellar parameters for white dwarfs. 

\section{Conclusions \label{conclusions}}

While traditional (and genuinely astrophysical) methods continue to be both useful and necessary to understand the new astronomical data, machine-learning techniques are needed to handle the sheer amounts of present and future data, and to lower the total CO$_2$ budget of astronomical research. 
Here we showed that both neural-network and tree-based algorithms, once sufficiently well trained, can be successfully employed to infer stellar parameters, distances, and extinctions in the absence of high-resolution spectroscopic data. In the near future, we will present a new {\tt StarHorse}-like catalogue for {\it Gaia} DR3 stars based on {\tt XGBoost} trained on high-resolution spectroscopic data.

%
%
\small  
%
\section*{Acknowledgments}   
%
This work was (partially) funded by the Spanish MICIN/AEI/10.13039/501100011033 and by ``ERDF A way of making Europe'' by the ``European Union'' through grant RTI2018-095076-B-C21, and the Institute of Cosmos Sciences University of Barcelona (ICCUB, Unidad de Excelencia 'Mar\'{\i}a de Maeztu') through grant CEX2019-000918-M. FA acknowledges financial support from MICINN (Spain) through the Juan de la Cierva-Incorporcion programme under contract IJC2019-04862-I. 

This work has made use of data from the European Space Agency (ESA)
mission {\it Gaia} (\url{http://www.cosmos.esa.int/gaia}), processed by the {\it Gaia} Data Processing and Analysis Consortium (DPAC,
\url{http://www.cosmos.esa.int/web/gaia/dpac/consortium}). Funding
for the DPAC has been provided by national institutions, in particular the institutions participating in the {\it Gaia} Multilateral Agreement.

%

%
\end{document}